\documentclass[twocolumn,preprintnumbers,prd,superscriptaddress,nofootinbib,floatfix,showpacs,showkeys]{revtex4-1}
\usepackage{amsmath}
\usepackage{amsfonts}
\usepackage{amssymb}
\usepackage{graphicx}
\usepackage{color}
\usepackage{hyperref}
\usepackage{booktabs}
\usepackage{hyperref}
\usepackage{cleveref}
\usepackage{braket}
\usepackage{mathtools}
\usepackage[rightcaption]{sidecap}
\usepackage{subfigure}
\usepackage{soul}
\usepackage{enumitem}
\usepackage{comment}
\usepackage{multirow}

\definecolor{vividviolet}{rgb}{0.62, 0.0, 1.0}
\definecolor{amaranth}{rgb}{0.9, 0.17, 0.31}
\definecolor{palatinateblue}{rgb}{0.15, 0.23, 0.89}
\definecolor{brightpink}{rgb}{1.0, 0.0, 0.5}
\definecolor{cornflowerblue}{rgb}{0.39, 0.58, 0.93}
\definecolor{deepcarminepink}{rgb}{0.94, 0.19, 0.22}
\definecolor{radicalred}{rgb}{1.0, 0.21, 0.37}
\hypersetup{ linktoc=all,
    colorlinks, linkcolor={palatinateblue},
    citecolor={brightpink}, urlcolor={amaranth}
}

\newcommand{\be}{\begin{equation}}
\newcommand{\ee}{\end{equation}}
\newcommand{\bs}{\begin{split}}
\newcommand{\bea}{\begin{eqnarray}}
\newcommand{\eea}{\end{eqnarray}}

\newcommand{\bes}{\begin{subequations}}
\newcommand{\ees}{\end{subequations}}

\begin{document}

\title{Impact of flavor condensate dark matter on accretion disk luminosity in spherical spacetimes}

\author{Antonio Capolupo}
\email{capolupo@sa.infn.it}
\affiliation{Dipartimento di Fisica ``E.R. Caianiello'' Universit\`{a} di Salerno, and INFN -- Gruppo Collegato di Salerno, Via Giovanni Paolo II, 132, 84084 Fisciano (SA), Italy}

\author{Orlando Luongo}
\email{orlando.luongo@unicam.it}
\affiliation{University of Camerino, Via Madonna delle Carceri, Camerino, 62032, Italy.}
\affiliation{Department of Nanoscale Science and Engineering, University at Albany-SUNY, Albany, New York 12222, USA.}
\affiliation{INAF - Osservatorio Astronomico di Brera, Milano, Italy.}
\affiliation{Istituto Nazionale di Fisica Nucleare (INFN), Sezione di Perugia, Perugia, 06123, Italy.}
\affiliation{Al-Farabi Kazakh National University, Al-Farabi av. 71, 050040 Almaty, Kazakhstan.}

\author{Aniello Quaranta}
\email{aniello.quaranta@unicam.it}
\affiliation{University of Camerino, Via Madonna delle Carceri, Camerino, 62032, Italy.}

\begin{abstract}
We investigate the impact of dark matter condensates on the emission and thermodynamic properties of accretion disks, in a spherically-symmetric and static background. We focus on a class of models where dark matter originates from a genuine mass mixing among neutrino fields and compute the corrections to the dark matter's potential within galactic halo. We find a corresponding Yukawa correction induced by the dark matter energy-momentum tensor over the Newtonian potential. In so doing, employing Schwarzschild coordinates, and adopting the Novikov–Thorne formalism, we compute the geodesic structure and the corresponding disk-integrated luminosity profiles. Assuming a constant mass accretion rate, constituted solely by baryonic matter, we find non-negligible deviations in both the disk structure and radiative output, as compared to the standard Schwarzschild case. Afterwards, we discuss physical consequences of our Yukawa correction, comparing it with recent literature, predicting similar potentials, albeit derived from extended theories of gravity. Accordingly, we thus speculate to use our results to distinguish among candidates of dark matter. Indeed, our findings suggest that incoming high-precision observations of accretion disk spectra may provide a tool to probe dark matter's nature under the form of particles, extended theories of gravity or condensates.
\end{abstract}

\maketitle
\tableofcontents

\section{Introduction}

The direct imaging of supermassive black hole shadows has indicated the existence of ultra-compact objects in galaxies \cite{Akiyama_2019,Pesce_2021, SMBH2}. These are both electromagnetically detectable and,  mainly, sources for gravitational radiation \cite{Cardoso2017,PhysRevLett.116.061102,10.1093/mnras/269.1.199},  providing \emph{de facto} new evidence for the emerging field of \emph{multi-messenger astrophysics} \cite{Mészáros2019}.

Despite observational progress, theoretical frameworks fail to predict the redshift-distribution of supermassive black holes, as well as their formation mechanisms and early growth phases. The latter ones remain poorly understood when compared to stellar black holes, puzzling the standard model of black hole formation \cite{Smith03042019,Zou_2024,10.1093/mnras/stae1819}.

In this respect, observational data indicate the presence of numerous supermassive black holes with masses exceeding $\sim10^9\,M_\odot$, placed at primordial times of the universe evolution, thus leading to a fundamental question: how could such massive compact objects form and grow to such scales within a so-limited interval of time?

To potentially answer this question, with the exception of Sgr $A^\star$ in the Milky Way and the supermassive black hole candidate in M87, \emph{the predominant method for estimating the mass of compact objects in galactic centers relies on the spectral analysis of their accretion disks} \cite{10.1093/mnras/266.1.137,10.1046/j.1365-8711.2002.05871.x,10.1093/mnras/stt157}.

Thus, the construction of accretion disk models appears quite relevant to clarify the nature of compact objects at all cosmic stages and,  necessarily, involves several simplifying assumptions, particularly concerning the kind of geometry determining the disk itself.

Significant results employ the use of the Kerr metric \cite{10.1111/j.1365-2966.2012.21074.x,Gammie_1998, Popham_1998, Abramowicz2013}, albeit departures are expected once the spacetime symmetry and properties appear different. Hence, beyond determining the right spacetime, it turns out to be crucial to formulate the correct background paradigm through which the accretion is modeled. To this end, we distinguish at least two main approaches, the \emph{Novikov-Thorne} \cite{NT1, NT2} and \emph{Bondi} \cite{EDGAR2004843,Ricotti_2007,PhysRevD.87.044007} descriptions around a gravity center, see e.g.  \cite{Abramowicz2013}.

In both of them, the usual strategy may be oversimplified, since it disregards potential deviations from the Kerr solution \cite{CHAUDHARY2025170006, Boshkayev_2022, Boshkayev2020, Kurmanov_2022}. To go through this problem, theoretical efforts have recently explored accretions in geometries that highly deviate from the standard Kerr line element, exhibiting accretion fluids with precise and alternative equations of state, see e.g. \cite{PhysRevD.106.063009, Bambi:2011vc, Boshkayev:2021chc, Uktamov:2024ckf,Tahelyani:2022uxw, Kurmanov:2025uwq,Cemeljic:2025bqz,Uniyal:2023inx,Sajadi:2023ybm, Kurmanov:2024hpn}.

These investigations are motivated in part by the fact that galaxies are embedded within extended \emph{dark matter halos}, whose gravitational influence dominates the dynamics at large galactocentric radii and may not be negligible even in proximity of the central object \cite{Boshkayev_2022,Boshkayev:2021chc,Boshkayev2020,Kurmanov_2022}. Further, these approaches seek possible departures from the \emph{Kerr hypothesis}, according to which compact objects might be solely characterized by two spacetimes, i.e., the Kerr and Schwarzschild metrics.

It is commonly plausible to consider that the central dark matter distribution may induce significant modifications to the local spacetime geometry \cite{CHAUDHARY2025170006, Boshkayev_2022, Boshkayev2020, Kurmanov_2022,PhysRevD.106.063009, Bambi:2011vc, Boshkayev:2021chc, Uktamov:2024ckf,Tahelyani:2022uxw, Kurmanov:2025uwq}.

In this regards, various theoretical models have been proposed to characterize the dark matter density profiles near galactic centers, as well as its potential interaction with compact objects \cite{10.1093/mnras/stz1698,PhysRevD.110.043034} and, in general, those depend on dark matter's nature. Examples of viable dark matter scenarios  lie on ultralight fields \cite{Ferreira2021,Rogers_2023,PhysRevD.95.043541}, hypermassive constituents \cite{PhysRevD.99.063015, Brandt_2016}, primordial black holes \cite{Green_2021}, solitons or particle-like configurations inside Einstein's theory \cite{Luongo:2025iqq,universe6120234} and so on\footnote{For a non-exhaustive review, refer to as Ref. \cite{universe5100213}.}.

In all the aforementioned cases, given the relativistic nature of the gravitational field close to supermassive black holes, it is reasonable to expect that the presence of a non-negligible dark matter component could alter the geodesic structure of the surrounding spacetime \cite{Boshkayev_2022,Boshkayev2020,Kurmanov_2022}. As a result, the dynamics of accretion flows, the formation of the innermost stable circular orbit (ISCO), and the spectral properties of the emitted radiation from the accretion disk may be significantly modified \cite{Boshkayev_2022,Boshkayev2020,Kurmanov_2022,CHAUDHARY2025170006}. These effects might be taken into account when interpreting observational data from compact sources embedded in dark matter envelopes \cite{Zavala:2019gpq}.

Motivated by the above considerations, here we explore the kinematic, thermodynamic, and spectral properties of accretion disks induced by \emph{dark matter under the form of condensates}. In particular, among all the possibilities reported above, dark matter is described here by a fermionic condensate, whose existence is constructed in analogy to standard condensates, induced by flavor mixing in the lepton sector \cite{3Flav, Capo2016,CosmoFlav,CurvFlav,Capolupo2025}, dubbed \emph{flavor condensate dark matter}  (FCDM). In this respect, geodesic motion and disk integrated luminosity, induced by dark matter condensates, within a spherically-symmetric configuration, are thus computed, emphasizing the main differences that one could expect with respect to the standard case, where dark matter does not exhibit condensate properties. In so doing, we work in isotropic coordinates, characterizing the spacetime structure, in which the energy-momentum tensor due to dark matter induces an exponential correction to the central Newtonian potential. Hence, as byproduct of dark matter's nature, we deal with the modification of the Newton's potential, rather than postulating a \emph{ad hoc} density distribution of the halo, yielding a more realistic scenario to characterize the presence of dark matter in spiral galaxies. As a consequence, by solving the Poisson equation, we infer an effective density distribution that is not fixed \emph{a priori}, but reobtained as consequence of the Yukawa potential inferred from dark matter condensate. The dark matter density distribution is thus compared with well-known density profiles and, accordingly, our main quantities are drawn within the Novikov-Thorne paradigm, showing the main departures from the standard Schwarzschild case,  previously discussed in the literature. Our numerical findings are thus reported, indicating the potential impact induced by dark matter condensate over the accretion formation. Consequently, once future measurements of the accretion disk luminosity are made, one can thus expect to distinguish \emph{a posteriori} the dark matter nature, by virtue of possible deviations induced by the condensate nature of our dark matter fluid.

The paper is organized as follows.
In Sect. \ref{sezione1}, we provide a brief overview of the flavor condensate dark matter and its fluid description.
In Sect. \ref{sezione2}, we report the geometrical background under exam. Afterwards, the overall kinematic analysis of our dark matter model is portrayed in Sect. \ref{sezione3}, showing the main departures over the geodesic equations. Accordingly, in Sect. \ref{sezione4}, we indicate our numerical findings, reporting the ranges of free parameters constrained within the most viable ranges. Finally, Sect. \ref{sezione5} points out conclusions and perspectives of this work.

\section{Flavor condensate dark matter and its gravitational analogues} \label{sezione1}

In the framework of quantum field theory with neutrino mixing in curved spacetime, the physical vacuum exhibits a nontrivial condensate structure, known as the \emph{flavor vacuum} $\ket{0_F(T)}$ \cite{CosmoFlav,CurvFlav,CurvNeut}.

It is possible to relate the ordinary vacuum, $\ket{0_M}$, to the above-mentioned $\ket{0_F(T)}$. To do so, we remark that the flavor vacuum is defined on a given hypersurface $\Sigma(T)$. In this picture, $T$ parametrizes spacetime foliation, i.e.,  $\mathcal{M} \simeq T \in \mathbb{R} \times \Sigma$, by the action of the mixing generator $\mathcal{J}_{\theta}(T)$, yielding so
\begin{equation}
    \ket{0_F(T)} = \mathcal{J}^{-1}_{\theta}(T) \ket{0_M}\ .
\end{equation}
Accordingly, the \emph{mixing generator} implies a rotation occurring from neutrino fields, with definite masses $\nu_1,\nu_2,\nu_3$, and  neutrino flavor fields, say $\nu_e, \nu_\mu,\nu_\tau$, by
\begin{equation}
    \nu_{\alpha}(T) = \mathcal{J}_{\theta}^{-1}(T) \nu_{j} \mathcal{J}_{\theta}(T) = \sum_{i=1,2,3} U_{\alpha i} \nu _i,
\end{equation}
having $(\alpha,j) = (e,1),(\mu,2),(\tau,3)$, $U_{\alpha i}$, the elements of the Pontecorvo-Maki-Nakagawa-Sakata matrix\footnote{Remarkably, the equalities hold on the hypersurface $T$. For additional details on the mixing transformation, see e.g. Refs.  \cite{CurvFlav,CurvNeut, Capolupo2025}.}  \cite{ParticleDataGroup:2024cfk}.

The flavor vacuum exhibits a condensate structure composed of neutrino–antineutrino pairs. Each pair exhibits definite mass eigenstates and, consequently, it possesses a non-trivial energy–momentum content.

In particular,  for a broad class of background geometries, among which the remarkable case of a spatially flat Friedmann–Robertson–Walker spacetime, and/or static, spherically symmetric metrics, the expectation value of the renormalized energy–momentum tensor evaluated on the flavor vacuum,
$\mathbb{T}_{\mu \nu}(T) = \bra{0_F(T)} T_{\mu \nu} \ket{0_F(T)}$, takes the form of a \emph{pressureless perfect fluid}, i.e., it is equivalent to the stress–energy tensor of a dust-like component\footnote{Similar results have been considered involving exotic generalization of K-essence models, see e.g. \cite{Luongo:2024opv}, implying a so-called quasi-quintessence field, investigated in Refs. \cite{DAgostino:2022fcx,Belfiglio:2023rxb}, with the impressive task of alleviating the cosmological constant problem \cite{Luongo:2018lgy,Belfiglio:2022qai,Luongo:2023jnb}.}
\begin{equation}\label{Fluid1}
    \mathbb{T}_{\mu\nu} \equiv \mathrm{diag} (\varepsilon,0,0,0).
\end{equation}

The case of umost interest concerns static and spherically symmetric configurations in the asymptotic regime, i.e., at sufficiently large distances from the gravitational center. This prerogative is explored in order to model galactic dark matter distributions within the aforementioned description, as described in what follows.

\subsection{The spherical case}

In the case of a static and spherically symmetric spacetime, in the weak field approximation, the Newtonian spacetime acquires the form,
\begin{equation}\label{newme}
ds_{EXT}^2 =(1+2V(r)) dt^2 - (1-2V(r))dr^2 -r^2d\theta^2 - r^2\sin^2 \theta d \phi^2,
\end{equation}
where the subscript ${\rm EXT}$ stands for the fact that the above metric is valid only \emph{far enough} from the central gravitational source, e.g. quite far from the event horizon of a black hole.

As reported explicitly in Appendix \ref{AppendixA}, the energy density turns out to be \cite{Capolupo2025}
\begin{equation}\label{Fluid2}
    \varepsilon (r) = \varepsilon_0(1+2V(r)),
\end{equation}
where the constant $\varepsilon_0$ is the flat space energy density of the flavor vacuum.

Its numerical value fixes the energy scale at which the condensate may exist. Accordingly, its numerical bound depends on an ultraviolet momentum cutoff,  $\Lambda$, namely $\varepsilon_0= \varepsilon_0 ( \Lambda)$ and, therefore, typical values lie on $\Lambda = \Lambda_{\mathrm{EW}} = 246 \ \mathrm{GeV}$, implying an electroweak scale, or alternatively employing gravitational mass, e.g. the baryonic mass $M_{G}$ of a galaxy, the mass $M_{BH}$ of a black hole, and so on.

Closely related to $\varepsilon_0$, one can formulate the corresponding \emph{Yukawa distance}, defined as
\begin{equation}\label{YukawaDistance}
    \delta(\Lambda) = \frac{1}{ \sqrt{8\pi \varepsilon_0(\Lambda)}} \ .
\end{equation}
Since $\varepsilon_0$ increseas with $\Lambda$, smaller cutoffs imply larger distances \cite{Capolupo2025}, intimately related to the corresponding Yukawa potential
\begin{equation}\label{FlavorPotentialX}
    V(r)=-\alpha m \frac{e^{-\frac{r}{\delta(\Lambda)}}}{r},
\end{equation}
allowing for describing the large-scale gravity within the metric considered above.

In Eq. \eqref{FlavorPotentialX}, $\alpha$ appears as a free  parameter, whereas $m$ is the  gravitational mass. Moreover, the gravitational Yukawa-like potential is specified through $\alpha$ and $\delta$, or equivalently by $\alpha$ and $ \Lambda$.

\begin{figure}
\includegraphics[width=\columnwidth]{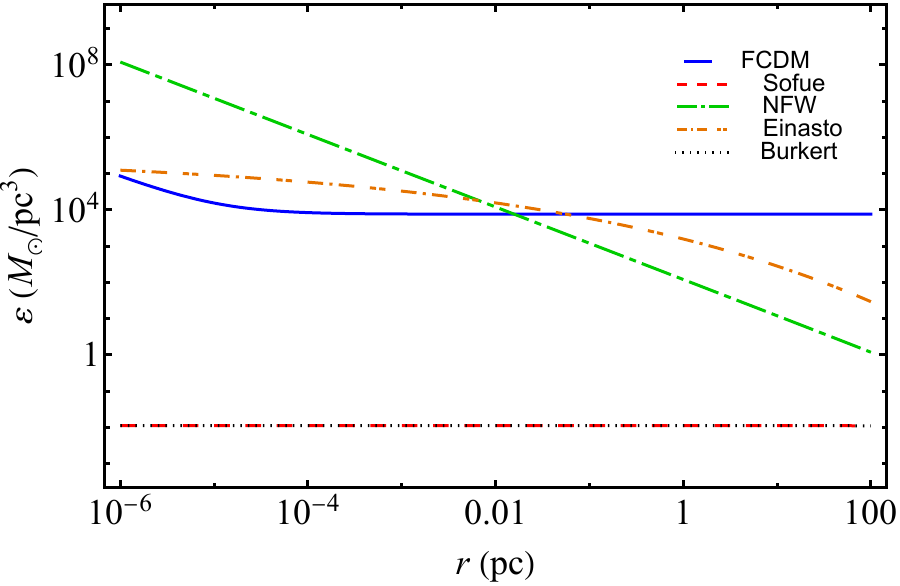}
    \caption{Logarithmic scale plot of the Energy density $\varepsilon(r)$ in $M_{\odot}/\mathrm{pc}^3$, as a function of $r$: for the FCDM profile \eqref{FCDMProfile} with $\alpha =-0.2$ and $\delta = 10 \ \mathrm{kpc}$; the Sofue profile $\varepsilon(r)=\varepsilon_{0,S} e^{-\frac{r
    }{r_0}}$ with  $\varepsilon_{0,S}=7.56 \times 10^{-3} \ M_{\odot}/\mathrm{pc}^3$; the NFW profile $\varepsilon(r)=\frac{\varepsilon_{0,NFW } \ r_0}{r\left(1+ r/r_0\right)^2}$ with $\varepsilon_{0,NFW } =  10^{-2} M_{\odot}/\mathrm{pc}^3$; the Burkert profile $\varepsilon(r)=\frac{\varepsilon_{0,B } \ r_0^3}{(r+r_0)\left(r^2 +r_0^2\right)}$ with $ \varepsilon_{0,B } = 11.1 \times 10^{-3} \ M_{\odot}/\mathrm{pc}^3$; the Einasto profile $\varepsilon(r) = \varepsilon_{0,E} e^{\frac{2\left(1-(r/r_0)^\sigma \right)}{\sigma} }$ with  $\varepsilon_{0,E}=20.1\times 10^{-3} \ M_{\odot}/\mathrm{pc}^3$ and $\sigma=0.12$. All the scale lengths are $r_0 = 12 \ \mathrm{kpc}$. We have employed the values reported in \cite{10.1093/mnras/stab2571} for the Milky Way.}
  \label{fig:Density}
\end{figure}

It has been shown that the potential correction of Eq. \eqref{FlavorPotentialX} may account for the missing gravitational matter in galaxies \cite{Capolupo2025}. We refer to the fluid described by Eqs. \eqref{Fluid1} and \eqref{Fluid2}, corresponding to the weak field potential of Eq. \eqref{FlavorPotentialX} as FCDM. By inserting \eqref{FlavorPotentialX} in Eq. \eqref{Fluid2}, we find the FCDM profile
\begin{equation}\label{FCDMProfile}
    \varepsilon(r) = \varepsilon_0\left(1-2\alpha m\frac{e^{-\frac{r}{\delta(\Lambda)}}}{r}\right)\ \ .
\end{equation}
A comparison with other dark matter density profiles is outlined in Fig.  \ref{fig:Density}.

The Yukawa correction to Newton's potential arises in a variety of other contexts, among which

\begin{itemize}
    \item[-] models with a massive graviton \cite{Inan_2024};
    \item[-] anti-gravity \cite{1984A&A...136L..21S}, where values in the range $-0.95<\alpha<-0.9$, $\delta \sim $ a few tens of $\mathrm{kpc}$;
    \item[-] the weak field limit of $F(R)$ modified theories of gravity \cite{CAPOZZIELLO2020100573}, where values of $\alpha \in [-1,1]$ are studied. It is worth mentioning also the specific case of $F(R) = R^n$ \cite{2007MNRAS.381.1103F}, where the gravitational potential is compatible with $\alpha = - \frac{1}{2}$ for $r \ll \delta $;
    \item[-] theories that generically modify gravity, e.g., in \cite{PhysRevD.104.043009} where a good agreement with the Milky way data is found in correspondence with negative $\alpha$ and $\delta$ scaling approximately as $\delta \propto 1/\sqrt{|\alpha|}$, with $\delta \simeq 0.77 \ \mathrm{kpc}$  for $|\alpha| = 1$. The value $\alpha \sim \pm  0.4$ and a scale dependent $\delta$ ($\delta \sim $ a few $\mathrm{kpc}$ on galactic scales) are found in \cite{Capolupo2025} for the Milky Way.
\end{itemize}
Despite the above analogies, the great advantage of the FCDM paradigm is to lie on general relativity only,  without any need of extending gravity as in the above cases.

Moreover, within the FCDM scheme, the Yukawa correction is ascribable to the presence of an additional energy density.

Accordingly, the main advantage is that the dark matter equation of state turns out to be exactly zero, resembling a dust behavior, in line with the hypothesis of non-relativistic dark matter constituent\footnote{For alternative equations of state, refer to as Refs. \cite{Luongo:2025iqq,Faber:2005xc}.}.

\section{Junctions on gravitational background}\label{sezione2}

The external spacetime, identified by the Newtonian metric above, describes the external structure of our solution, mimicking the halo of dark matter. The mass generator, i.e., the gravitational charge acting as source might be described and matched with exterior.

To do so, we consider a static, non-rotating and isotropic spacetime, that in the Schwarzschild chart $(t,r,\theta,\phi)$ corresponds to the metric
\begin{equation}\label{GeneralMetric}
    ds^2 =f(r)dt^2-g(r)dr^2 -r^2d\theta^2 -r^2\sin^2\theta d\phi^2 \ ,
\end{equation}
where $f,g$ are sufficiently regular functions of the radius $r$. We introduce a separation radius $r_b$, a free parameter of the model that may be understood as the inner radius of the dark matter envelope \cite{Boshkayev2020}. For $r\leq r_b$ we assume the Schwarzschild black hole solution, corresponding to
\begin{equation}\label{SchwarzschildSolution}
    f_0(r)=1-\frac{2M_{BH}}{r}, \ \ \ \ \ \ \ \ \ g_0(r)=\left(1-\frac{2M_{BH}}{r}\right)^{-1} \ .
\end{equation}
The separation radius $r_b$ is assumed to be large enough that for $r>r_b$ the effect of the FCDM envelope is accurately reproduced by the weak field potential of Eq. \eqref{FlavorPotentialX}. For radii above $r_b$ the metric is perturbed by the FCDM envelope, with weak field potential $V(r)$ provided in Eq. \eqref{FlavorPotentialX}. The metric functions beyond $r_b$ are then
\begin{equation}\label{DMSolution}
f(r)=f_0(r)+2V(r); \ \ \ \ \ \ \ g(r) = g_0(r)-2V(r) \ ,
\end{equation}
i.e. the Schwarzschild metric perturbed by the weak field due to the FCDM envelope\footnote{To be precise one should add to $V$ the overall potential due to the baryonic matter $V_{B}$. While $V_{B}$ is fundamental for the correct determination of the galactic properties (e.g. the rotation curves, on scales of a few $\mathrm{kpc}$), we expect it to be negligible over the scales relevant for accretion $\sim$ a few tens of $\mathrm{au}$. In other words, we can assume that the enclosed baryonic mass is negligible with respect to the black hole mass $M_B(r \sim 10-100 \ \mathrm{au}) \ll M_{BH}$, and with respect to the dark matter envelope.}.

In order to ensure continuity at $r=r_b$ we shift the FCDM potential by a constant
\begin{equation}
    \tilde{V}(r) = V(r)-V(r_b) \ .
\end{equation}
and we set
\begin{equation}
    f(r)=\begin{cases} f_0(r) & r\leq r_b \\
    f_0(r) + 2\tilde{V}(r) & r> r_b
    \end{cases} \ , \label{fCompl}
\end{equation}
\begin{equation}
     g(r)=\begin{cases} g_0(r) & r\leq r_b \\
    g_0(r) - 2\tilde{V}(r) & r> r_b
    \end{cases} \ . \label{gCompl}
\end{equation}
Notice that the derivatives of the metric functions are discontinuous at $r_b$. We remark that the perturbative treatment adopted here is meaningful as long as $|\tilde{V}(r)| \ll f_0(r), g_0(r).$ In practice this condition is verified already for $r\geq r_b = 6M_{BH}$. In Eqs. \eqref{fCompl} and \eqref{gCompl}, $f_0, g_0$ are geometric consequences of the gravitational charge that determines the accretion, i.e., the mass determined by the Schwarzschild solution, while $\tilde{V}(r)$ includes the effect of the dark matter halo.

Remarkably, for large enough radii, namely  $r\geq r_b \sim 6 M_{BH}$, even the Schwarzschild solution may be approximated in a  weak field regime by $f_0(r) = 1 - \frac{2M_{BH}}{r}, g_0\simeq 1 + \frac{2M_{BH}}{r}$, allowing us to write
\begin{equation}
    f(r) = 1 + \frac{2m(r)}{r} ;\ \ \ g(r) \simeq \left(1- \frac{2m(r)}{r}\right)^{-1}
\end{equation}
yielding an effective mass,
\begin{equation}
m(r) =    \begin{cases}
        M_{BH} & r \leq r_b \\
        M_{BH}\left(1+\alpha e^{-\frac{r}{\delta }}\right) & r> r_b
    \end{cases} \ .
\end{equation}

\section{Geodesic Motion and Disk Luminosity}\label{sezione3}

To determine the thin accretion disk properties, we focus on the geodesic motion close to the equatorial plane $\theta \rightarrow \frac{\pi
}{2}$, where each coefficient of Eq.  \eqref{GeneralMetric} depends only on $r$. For circular orbits one can derive the angular velocity, $\Omega$, the specific energy, $E$, and the specific angular momentum, $L$, of test particles  \cite{Harko2009,Boshkayev2020}. From Eq. \eqref{GeneralMetric}, we have

\begin{subequations}
    \begin{align}
    &\Omega = \frac{d\phi}{dt} =\sqrt{-\frac{
    \partial_rg_{tt}}{\partial_r g_{\phi\phi}}} = \sqrt{\frac{f'(r)}{2r}} ,\label{AngularVelocity}\\
    &E = g_{tt} \frac{dt}{d\tau}=\frac{g_{tt}}{\sqrt{g_{tt}+ \Omega^2 g_{\phi\phi}}}=\frac{f(r)}{\sqrt{f(r)-r^2 \Omega^2}},\label{SpecificEnergy}\\
    &L =-g_{\phi \phi}\frac{d\phi}{d\tau}=-\frac{\Omega g_{\phi \phi}}{\sqrt{g_{tt}+ \Omega^2 g_{\phi\phi}}}=\frac{r^2 \Omega}{\sqrt{f(r)-r^2 \Omega^2}}.\label{SpecificAngularMomentum}
    \end{align}
\end{subequations}
Above, the prime denotes derivative with respect to $r$, whereas $\tau$ denotes the proper time.

It is customary to define dimensionless quantities, reported in geometrized units, from Eqs. \eqref{AngularVelocity}, \eqref{SpecificEnergy}
and \eqref{SpecificAngularMomentum}, with reference to a typical mass. Hence, we define $\Omega^*=M_{BH}\Omega$, $L^*=\frac{L}{M_{BH}}$ and $E^* = E$.
In the Novikov-Thorne model of accretion \cite{NT1,NT2}, these quantities determine the radiative flux emitted from the disk
\begin{equation}\label{Radiative flux}
    \mathcal{F}(r) = -\frac{\dot{M}_{BH}}{4\pi r \sqrt{g(r)f(r)}} \frac{\Omega'(r)}{(E-\Omega L)^2} \int_{r_i}^{r} d\rho (E-\Omega L) L'(\rho) \ .
\end{equation}
Accordingly,
\begin{itemize}
    \item[-] The mass accretion rate of the black hole $\dot{M}_{BH}$ is determined by the amount of mass-energy per unit time falling inwards.
    \item[-] We assume that $\dot{M}_{BH}$ is exclusively due to ordinary matter, such as baryonic, gas, dust, etc., falling in the black hole and we further take it to be constant for simplicity.
\end{itemize}
On the one hand, this ensures that the FCDM profile remains static during accretion and, on the other hand, it is consistent with the condensate nature of the FCDM, conversely to particle dark matter, which could instead contribute to the accretion flow.

The lower integration bound, $r_i$, is the disk ISCO, defined by $L'(r_i)=0$. Conventionally introducing a rescaled flux as $\mathcal{F}^*(r) = M_{BH}^2 \mathcal{F}(r)$, from Eq. \eqref{Radiative flux} one can derive the luminosity for an observer at infinity $\mathcal{L}_{\infty}$ as follows \cite{Boshkayev2020,NT1,NT2}. First, the differential of $\mathcal{L}_{\infty}$ is related to the flux through
\begin{equation}\label{Flux-Luminosity}
   \frac{d \mathcal{L}_{\infty}}{d \ln r} = 4\pi r  \sqrt{f(r)g(r)}E(r) \mathcal{F}(r) \ .
\end{equation}
Afterwards, by comparing the flux with the Stefan-Boltzmann law, one defines a characteristic temperature

\begin{equation}
T_{*} = \frac{\dot{M}_{BH}}{4 \pi \sigma M_{BH}^2},
\end{equation}
where $\sigma$ is the Stefan-Boltzmann constant. Finally, assuming a blackbody emission from the disk and taking into account the gravitational redshift $z =\frac{dt}{d\tau}-1$, one obtains the spectral luminosity at infinity as
\begin{eqnarray}\label{Spectral Luminosity}
\nonumber \nu \mathcal{L}_{\nu,\infty} &=& \frac{60}{\pi^3 M_{BH}^2} \int_{r_i}^{\infty} dr  \frac{r\sqrt{f(r)g(r)}E(r)\tilde{y}^4(r)} {\exp{\left[\tilde{y}(r)/{\mathcal{F}^{*\frac{1}{4}}(r)}\right]}-1} ,
\end{eqnarray}
where we have introduced the dimensionless variables $y= \frac{h\nu}{k_BT_{*}}$, $\tilde{y}(r) = y(1+z(r))$, $h$ and $k_B$ being respectively Planck's and Boltzmann's constant.

These quantities can be used to compute our outcomes, referring to a dark matter envelope induced by our model. This prerogative will be discussed in what follows.

\section{Numerical results}\label{sezione4}

We now perform a numerical analysis of the quantities relevant to accretion, as introduced in the previous section. We start by involving viable values for the black hole mass, radius and rate, as

\begin{subequations}
    \begin{align}
    &M_{BH} = 4.993 \ \mathrm{au},\\
    &r_b = 6 M_{BH},\\
    &\dot{M}_{BH}=10^5,
    \end{align}
\end{subequations}
where the mass accretion rate is dimensionless, in geometric units.

Note that, even though the value of $r_b$ is picked arbitrarily, it ensures the  weak field approximation to hold, as in Eqs. \eqref{fCompl} and \eqref{gCompl}.

The Yukawa parameters $\alpha,\delta$ are only partially constrained, and are expected to scale with the size of the gravitating system \cite{CAPOZZIELLO2020100573,Capolupo2025,PhysRevD.104.043009}. The parameter $\alpha$ usually spans within the range $\left[-1,1\right]$, i.e., a positive value of $\alpha \simeq 0.4$, along with $\delta \simeq 1 \ \mathrm{kpc}$, provides a good agreement with the rotation curves of the Milky Way, see e.g. Ref.  \cite{Capolupo2025}. Conversely, negative values appear favored by other analyses, see e.g. \cite{1984A&A...136L..21S, PhysRevD.104.043009}, with an interesting extreme value,  $\alpha=-0.92$ found in Ref. \cite{1984A&A...136L..21S}.

Constraints on the fundamental plane of elliptical galaxies \cite{CAPOZZIELLO2020100573} favor relatively smaller values of $|\alpha| \leq 0.2$, whereas larger negative values of $\alpha$ are contemplated in the Milky Way analysis of Ref.  \cite{PhysRevD.104.043009}, along with an appropriate scaling of\footnote{According to the analysis performed in Ref.  \cite{PhysRevD.104.043009},  $\delta = \eta |\alpha|^c$, where $\eta = 0.77 \ \mathrm{kpc}$ and $c \simeq -0.503$.} $\delta$.

In addition to the scaling of $\delta$ with the size of the gravitating system, $\delta$ and $\alpha$ appear to be mildly correlated, with lower values of $\delta$ leading to lower values of $\alpha$ \cite{PhysRevD.104.043009, Capolupo2025}.

For our numerical analysis, we work out different values of $\alpha$ and $\delta$, in order to cover all the ranges discussed in the literature. Thus,
\begin{enumerate}
    \item $\alpha = 0$. Here, the metric in Eq.  \eqref{GeneralMetric} reduces to Schwarzschild everywhere, yielding ordinary accretion around a Schwarzschild black hole, having an ISCO as $r_{ISCO} = r_{ISCO,0}= 6 M_{BH} = 29.598 \ \mathrm{au}$,  for our chosen mass.
    \item $\alpha = -0.92$, $\Lambda = \Lambda_1 = 10 \ \mathrm{GeV}  \rightarrow \delta = \delta_1  = 333 \ \mathrm{au}$. This value of $\alpha$ is that of Ref. \cite{1984A&A...136L..21S}. The cutoff $\Lambda$ is chosen to reproduce $\delta$ of the order of a few hundreds of $\mathrm{au}$. The ISCO is $r_{ISCO,1} = 63.085 \ \mathrm{au}$.
    \item $\alpha = -0.2$, $\Lambda = \Lambda_2=1.6 \ \mathrm{keV} \rightarrow \delta = \delta_2 = 10 \ \mathrm{kpc}$. These values are compatible with those analyzed in Ref. \cite{CAPOZZIELLO2020100573}. The ISCO is $r_{ISCO,2} = 25.370 \ \mathrm{au}$.
    \item $\alpha = 0.8$, $\Lambda = \Lambda_3=1.6 \ \mathrm{keV} \rightarrow \delta = \delta_3 = 10 \ \mathrm{kpc}$. These values are compatible with those analyzed in \cite{CAPOZZIELLO2020100573}. The ISCO is $r_{ISCO,3} = 42.06 \ \mathrm{au}$.
    \item $\alpha = -0.98$, $\Lambda = \Lambda_4=4 \ \mathrm{GeV} \rightarrow \delta = \delta_4 = 832 \ \mathrm{au}$. The value of $\alpha$ is determined from is a rather extreme negative value, compatible with the values discussed in \cite{PhysRevD.104.043009}, whereas $\delta$ is chosen to be around $10^3 \ \mathrm{au}$ as an intermediate value between the disk region (a few $\mathrm{au}$) and the galactic scale ($\simeq \mathrm{kpc}$). The ISCO is $r_{ISCO,4} = 142.183 \ \mathrm{au}$.
    \item $\alpha = 0.4$, $\Lambda = \Lambda_5= 15.425 \ \mathrm{keV} \rightarrow \delta = \delta_5 = 1.06 \ \mathrm{kpc}$. These values are obtained from the analysis of the rotation curves of galaxies \cite{Capolupo2025} and are likewise compatible with the analysis presented in \cite{PhysRevD.104.043009}. The ISCO is $r_{ISCO,5} = 36.562 \ \mathrm{au}$
\end{enumerate}

\begin{figure}
\includegraphics[width=\columnwidth]{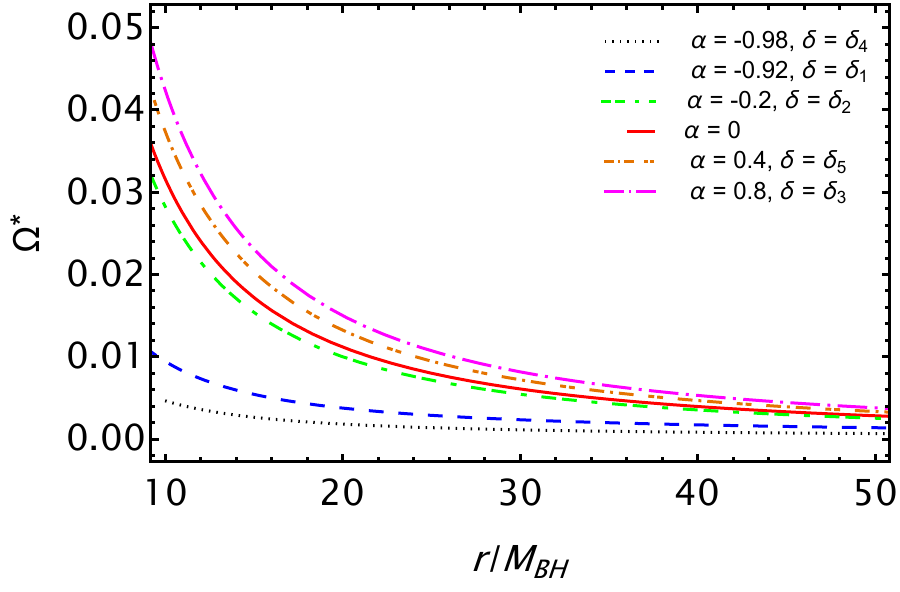}
    \caption{Rescaled angular velocity $\Omega^*=M_{BH} \Omega$, from Eq. \eqref{AngularVelocity} for different values of the parameters $\alpha,\delta$, as a function of $r$. $\alpha = 0$ corresponds to the Schwarzschild metric.}
    \label{fig:AngularVelocity}
\end{figure}

\begin{figure}
\includegraphics[width=\columnwidth]{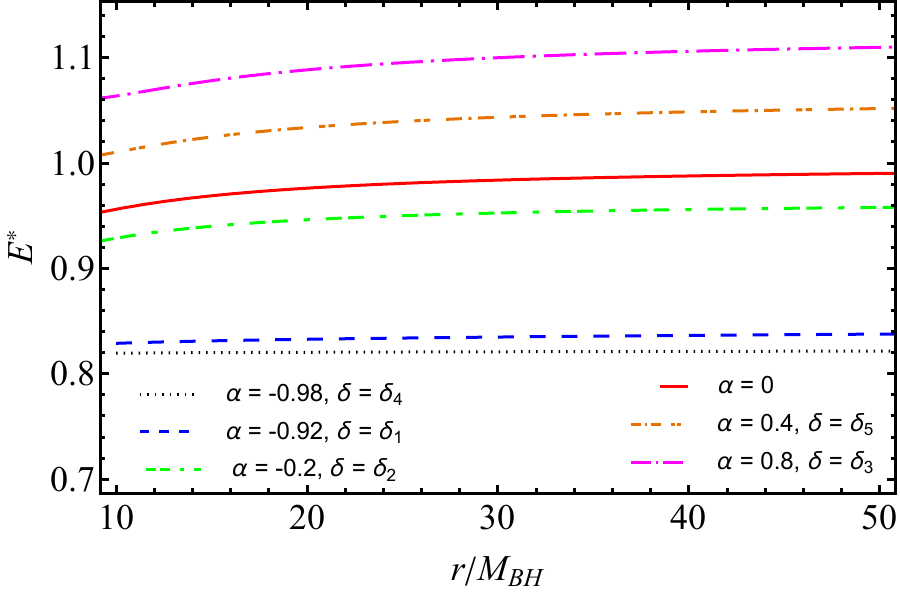}
    \caption{Specific energy $E^*=E $, from Eq. \eqref{SpecificEnergy} for different values of the parameters $\alpha,\delta$, as a function of $r$. $\alpha = 0$ corresponds to the Schwarzschild metric.}
    \label{fig:SpecificEnergy}
\end{figure}

\begin{figure}
\includegraphics[width=\columnwidth]{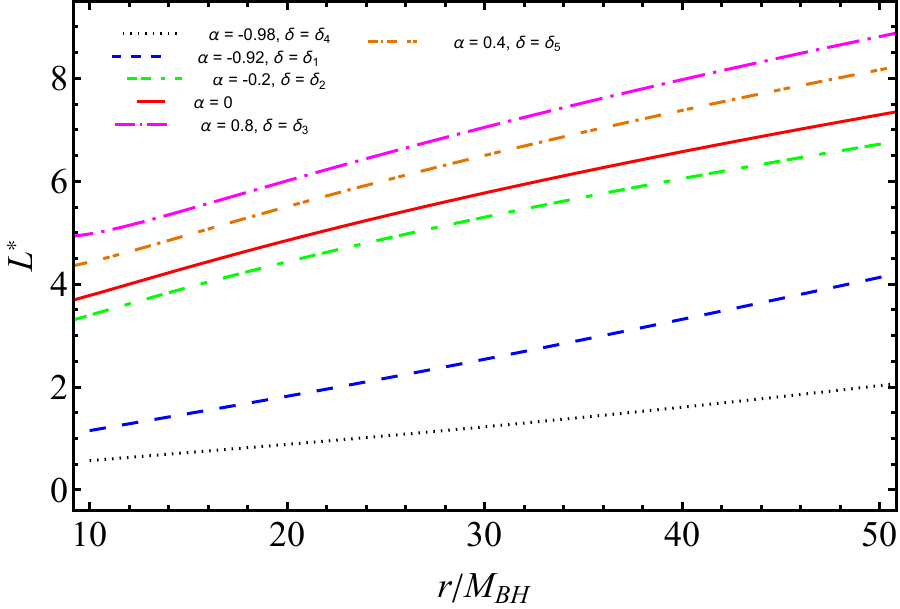}
    \caption{Specific angular momentum $L^*=L/M_{BH} $, from Eq. \eqref{SpecificAngularMomentum} for different values of the parameters $\alpha,\delta$, as a function of $r$. $\alpha = 0$ corresponds to the Schwarzschild metric.}
    \label{fig:SpecificAngularMomentum}
\end{figure}
In Fig. \ref{fig:AngularVelocity} we plot the angular velocity as a function of $r$. The effect of the FCDM potential \eqref{FlavorPotentialX} is more pronounced at smaller radii and shows a marked dependence on $\alpha$.

\begin{figure}
\includegraphics[width=\columnwidth]{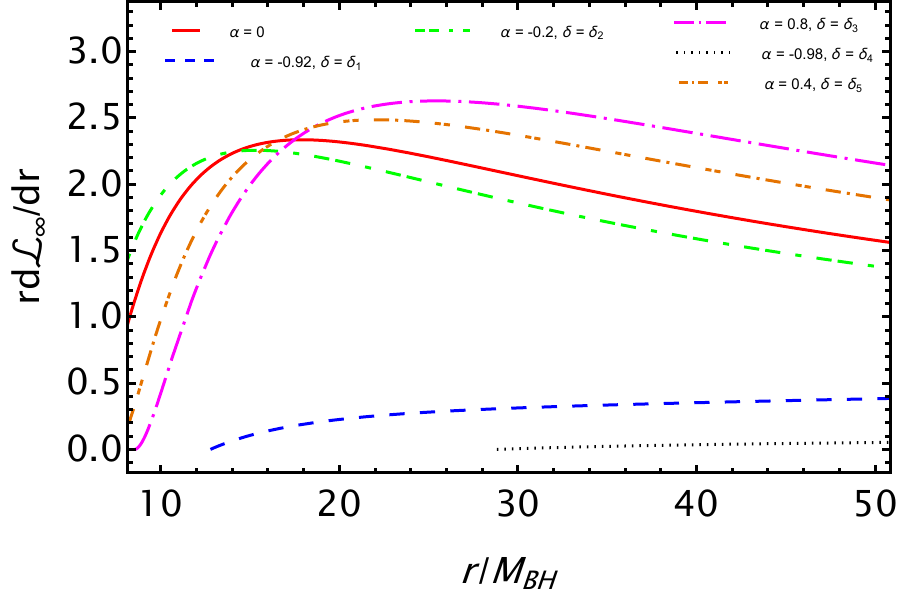}
    \caption{Differential Luminosity $\frac{d\mathcal{L_{\infty}}}{d \ln r} $, from Eq. \eqref{Flux-Luminosity} for different values of the parameters $\alpha,\delta$, as a function of $r$. $\alpha = 0$ corresponds to the Schwarzschild metric.}
\label{fig:DiffLuminosity}
\end{figure}

Larger positive and negative values of $\alpha$ visibly impact the specific energy (Fig. \ref{fig:SpecificEnergy}) in the range considered $r \in [10,50] M_{BH}$. Depending on the value of $\delta$, the specific energy quickly converges  to the Schwarzschild value at larger $r$. Similar conclusions are valid also for the specific angular momentum, plotted in Fig. \ref{fig:SpecificAngularMomentum}.
\begin{figure}
	\includegraphics[width=\columnwidth]{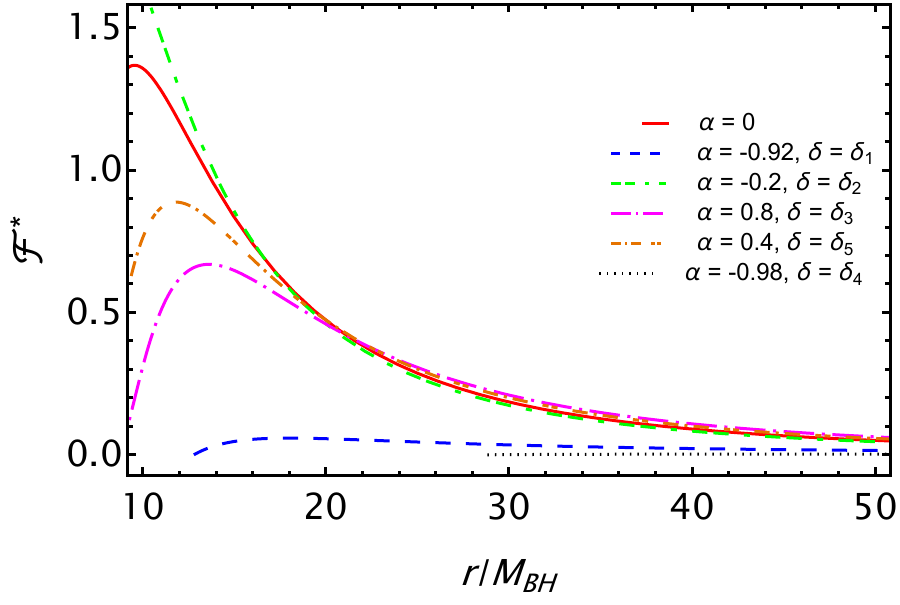}
    \caption{Radiative Flux $\mathcal{F}^*=\mathcal{F}M^2_{BH} $, from Eq. \eqref{Radiative flux} for different values of the parameters $\alpha,\delta$, as a function of $t$. $\alpha = 0$ corresponds to the Schwarzschild metric. }
    \label{fig:RadiativeFlux}
\end{figure}
At radii $ 10 \ M_{BH}< r < 30 \ M_{BH}$ the radiative flux, as shown in Fig. \ref{fig:RadiativeFlux}, is significantly reduced for almost all the values of $\alpha$ considered. The only exception is $\alpha=-0.2, \delta =\delta_2 = 10 \ \mathrm{kpc}$, for which the radiative flux is slightly enhanced at small radii. This is immediately understood in terms of the location of the ISCO for such parameters, i.e. $r_{ISCO,2} = 25. 370 \ \mathrm{au} < r_{ISCO,0}$. Since the $ISCO$ is shifted inwards, the lower integration bound in \eqref{Radiative flux} is lowered, increasing $\mathcal{F}(r)$ at small radii. For all the other choices of parameters the $ISCO$ is instead shifted outwards, leading to a reduced flux.
The reduction is more pronounced at larger positive and negative values of $\alpha$. For the large negative values $\alpha = -0.98, -0.92$, the flux in the regions $r < 30  \ M_{BH}, r< 10 \ M_{BH}$ vanishes. The ISCO is,  indeed, shifted towards larger radii, respectively beyond $30 \ M_{BH}$ and $10 \ M_{BH}$. The shift of the ISCO and the consequent impact on the disk emission is visible also in the differential luminosity of Fig. \ref{fig:DiffLuminosity}. The differential luminosity is indeed seen to vanish for the large negative values $\alpha = -0.98, -0.92$, in the regions $r < 30  \ M_{BH}, r< 10 \ M_{BH}$. All the values considered, except $\alpha=-0.2, \delta =\delta_2 = 10 \ \mathrm{kpc}$, yield a reduced luminosity at radii $r < 15 \ M_{BH}$, as compared to the Schwarzschild case.
\begin{figure}
\includegraphics[width=\columnwidth]{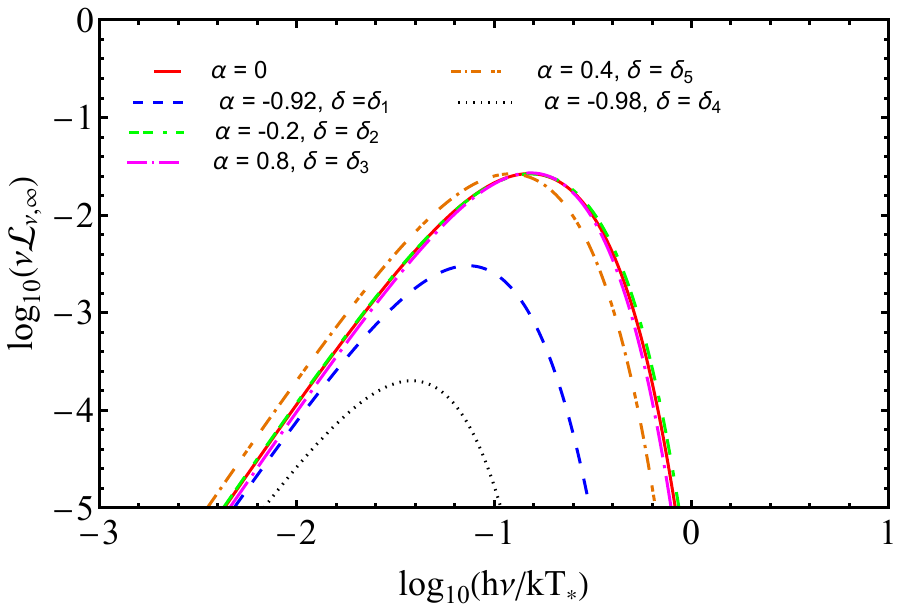}
    \caption{Logarithmic scale plot of the spectral Luminosity $\nu \mathcal{L_{\nu,\infty}} $, from Eq. \eqref{Spectral Luminosity} for different values of the parameters $\alpha,\delta$, as a function of $y = \frac{h\nu}{k_BT_{*}}$. $\alpha = 0$ corresponds to the Schwarzschild metric.}
\label{fig:SpectralLuminosity}
\end{figure}

In general, the presence of the FCDM leads to reduced luminosity, particularly at higher frequency values, as shown in Fig.  \ref{fig:SpectralLuminosity}. The phenomenology portrayed in the above plots has a rather simple interpretation when $\delta $ is roughly $\sim\mathrm{kpc}$. When $r\gg M_{BH}$, the metric \eqref{GeneralMetric} assumes the weak field form, with potential $V= -\frac{M_{BH}}{r}\left(1+\alpha e^{-\frac{r}{\delta}}\right)$. If $r\ll \delta$, the exponential approaches unity, yielding the potential $V \simeq -\frac{M_{BH}(1+\alpha)}{r}$. In this regime, the FCDM effectively modifies the value of the black hole mass by a factor $(1+\alpha)$, which in turn implies a shift of the ISCO by the same factor, ultimately affecting the radiative flux, in Eq.  \eqref{Radiative flux}.
This intuitive picture breaks down when $\delta$ is comparable to the accretion scales $\delta \simeq 10 - 100 \ \mathrm{au}$. In this case the exponential is significantly different from unity, and the effect of the Yukawa term is more involved. The ISCO is generally shifted outwards and the gravitational redshift $z(r)$ is enhanced, leading to a decreased luminosity.
\section{Final remarks}\label{sezione5}

In this work, we investigated the kinematic, thermodynamic, and spectral properties of accretion disks influenced by a dark matter condensate that contributes to the overall halo.

Specifically, we considered dark matter composed of \emph{fermionic condensates}, analogously originating from mass mixing mechanisms \cite{3Flav,Capo2016,CosmoFlav,CurvFlav,CurvNeut}, in close analogy to standard condensate formation.

To do so, we emphasized the formal mathematical structure that induces the presence of fermion condensates and, subsequently, wrote the corresponding basic assumptions, made in weak field regime. Accordingly, our final output corresponds to a deviation of Newton's potential that agrees with a \emph{Yukawa-like contribution}, similar to what found in frameworks that extend or modify gravity \cite{CAPOZZIELLO2020100573}.

Within this picture, we computed the deviations found within  geodesic motion and the corresponding disk-integrated luminosity, arising from the presence of such condensates, assuming a spherically symmetric configuration.

To this end, we employed Schwarzschild coordinates to characterize the underlying spacetime geometry, wherein the dark matter energy-momentum tensor induces an exponential correction to the central Newtonian potential.

Junction conditions between interior and exterior setups are also explored and motivated by viable bounds, that agree with previous literature.

All relevant observables were thus derived within the Novikov-Thorne formalism, which enabled a direct comparison with the standard Schwarzschild scenario and highlighted the key departures induced by the condensate dark sector. In this respect, ensuring that our dark matter envelope is static, we went through a constant mass rate to compute all our observable quantities, such as luminosity, temperature, and so on.

Our numerical analysis revealed non-negligible deviations in the disk structure and luminosity profile attributable to the dark matter condensate. These deviations suggested that, upon future precise measurements of accretion disk emission spectra, it may be possible to discriminate \emph{a posteriori} the fundamental nature of dark matter by identifying specific imprints left by its condensate origin.

Remarkably, a direct comparison with the standard Schwarzschild case has been developed, quantifying how the curves deviate from this standard case. Accordingly, we enlarged the ranges of parameters as much as possible to figure the departures out and interpreted this point as a plausible observable signature of condensate dark matter.

As perspectives, we will explore more deeply the mass mixing in alternative dark matter scenarios, employing, for example, other typologies of fields. In addition, we will explore additional models of accretion to distinguish whether the background could or not influence the final findings that we do expect. In this regard, we will also investigate additional spacetimes in which we analyze our dark matter distribution and consequent luminosity and thermodynamic properties.

\section*{Acknowledgements}
OL and AQ acknowledge the support by the  Fondazione  ICSC, Spoke 3 Astrophysics and Cosmos Observations. National Recovery and Resilience Plan (Piano Nazionale di Ripresa e Resilienza, PNRR) Project ID $CN00000013$ ``Italian Research Center on  High-Performance Computing, Big Data and Quantum Computing" funded by MUR Missione 4 Componente 2 Investimento 1.4: Potenziamento strutture di ricerca e creazione di ``campioni nazionali di R\&S (M4C2-19)" - Next Generation EU (NGEU).

\appendix

\section{Flavor condensate energy density in the Schwarzschild chart}\label{AppendixA}

In this appendix we derive the energy density due to the flavor condensate in the Schwarzschild chart $(t,r,\theta,\phi)$. We will follow steps analogous to those performed in the original derivation in isotropic coordinates, see Ref.  \cite{Capolupo2025}. In the weak field metric, Eq. \eqref{newme},
the energy density due to the flavor vacuum may be expanded as
\begin{equation}
    \varepsilon(r) = \varepsilon_0 + \varepsilon_1 V(r) + O(V^2(r))
\end{equation}
where $\varepsilon_0$ is the (constant) flat space energy density and terms of higher order in the potential can be neglected since $V(r) \ll 1$ by assumption. In order to extract $\varepsilon_1$ we proceed as follows. Let $\Psi_{0,J} (r)$ and $\Phi_{0,J}(r)$ be the solutions to the radial part of the Dirac equation in flat space, with $J$ denoting the collection of quantum numbers $J\equiv \lbrace E,j,\kappa,m_j\rbrace$. These are normalized according to
\begin{equation}\label{Normalization1}
    \int_0^{\infty}dr \ r^2 (\Psi_{0,J}\Psi_{0,J'}+\Phi_{0,J}\Phi_{0,J'}) = \delta_{J,J'}.
\end{equation}
To obtain the modes in the weak field metric, we rescale the flat space solutions according to
\begin{equation}
    \Psi_J = \left(1+ \frac{\gamma}{2}V(r)\right) \Psi_{0,J}; \ \ \ \Phi_J = \left(1+ \frac{\gamma}{2}V(r)\right) \Phi_{0,J},
\end{equation}
and impose the same normalization of Eq. \eqref{Normalization1}. Up to linear order in the potential, the normalization condition reads
\begin{eqnarray}
   \nonumber \int_0^{\infty} dr \ r^2  (1-V(r))(\Psi_{J}\Psi_{J'}+\Phi_{J}\Phi_{J'}) = \delta_{J,J'} \ \Rightarrow \\ \nonumber
   \int_0^{\infty}dr \ r^2 (1-V)\left(1+\gamma V + O(V^2)\right)(\Psi_{0,J}\Psi_{0,J'}+\Phi_{0,J}\Phi_{0,J'}) \\
   \nonumber \simeq \int_0^{\infty} dr \ r^2 \left(1+ \left(\gamma-1\right)V \right)(\Psi_{0,J}\Psi_{0,J'}+\Phi_{0,J}\Phi_{0,J'}) = \delta_{J,J'} \ .
\end{eqnarray}
Comparing with Eq. \eqref{Normalization1} we deduce $\gamma=1$. The energy density has the form \cite{Capolupo2025}
\begin{equation}
\varepsilon =\sqrt{1+2V(r)} \sum_{J} E   |W_J|^2 \left(\Psi^2_J + \Phi_j^2 \right),
\end{equation}
where the generalized sum $\sum_{J}$ involves an integral over the energies $E$ and $W_J$ is a coefficient independent of $r$ and $V$. Expanding up to linear order in $V$, we find
\begin{equation}
    \varepsilon = \varepsilon_0(1+V(r) + \gamma V(r)) = \varepsilon_0 (1+2V(r))\ ,
\end{equation}
implying $\varepsilon_1 = 2 \varepsilon_0$. In order to determine $V$ we now solve the Poisson equation
\begin{eqnarray}
    \nonumber \frac{1}{r^2} \partial_r (r^2 \partial_rV) = 4 \pi \varepsilon(r) \ \Rightarrow \\
    V'' + \frac{
    2
    }{r}V'-4\pi \varepsilon_0(1+2V)=0 \ .
\end{eqnarray}
Introducing $U(r) = r\left(1+2V(r)\right)$ and $\delta = \frac{1}{\sqrt{8\pi \varepsilon_0}}$ the equation simplifies to $
    U''-\frac{1}{\delta^2}U=0$.

The positive root must be discarded to ensure $V(r \rightarrow \infty) \rightarrow0$. Up to an additive constant we therefore end up with
\begin{equation}\label{YukawaPotential}
  V(r) = C \frac{e^{-\frac{r}{\delta}}}{r} \ ,
\end{equation}
with $C$ a multiplicative integration constant. When \eqref{YukawaPotential} is added to the Newtonian potential generated by a mass $m$, it is customary to write $C=-\alpha m$.


\begin{thebibliography}{10}

\bibitem{Akiyama_2019}
The Event Horizon~Telescope Collaboration and K.~et~al. Akiyama.
\newblock First m87 event horizon telescope results. i. the shadow of the supermassive black hole.
\newblock {\em The Astrophysical Journal Letters}, 875(1):L1, apr 2019.

\bibitem{Pesce_2021}
Dominic~W. Pesce, Daniel C.~M. Palumbo, Ramesh Narayan, Lindy Blackburn, Sheperd~S. Doeleman, Michael~D. Johnson, Chung-Pei Ma, Neil~M. Nagar, Priyamvada Natarajan, and Angelo Ricarte.
\newblock Toward determining the number of observable supermassive black hole shadows.
\newblock {\em The Astrophysical Journal}, 923(2):260, dec 2021.

\bibitem{SMBH2}
{The Event Horizon Telescope Collaboration} and {Akiyama, Kazunori} et~al.
\newblock The persistent shadow of the supermassive black hole of m87 - i. observations, calibration, imaging, and analysis.
\newblock {\em A\&A}, 681:A79, 2024.

\bibitem{Cardoso2017}
Vitor Cardoso and Paolo Pani.
\newblock Tests for the existence of black holes through gravitational wave echoes.
\newblock {\em Nature Astronomy}, 1(9):586--591, Sep 2017.

\bibitem{PhysRevLett.116.061102}
B.~P. et~al. Abbott.
\newblock Observation of gravitational waves from a binary black hole merger.
\newblock {\em Phys. Rev. Lett.}, 116:061102, Feb 2016.

\bibitem{10.1093/mnras/269.1.199}
Martin~G. Haehnelt.
\newblock Low-frequency gravitational waves from supermassive black holes.
\newblock {\em Monthly Notices of the Royal Astronomical Society}, 269(1):199--208, 07 1994.

\bibitem{Mészáros2019}
P{\'e}ter M{\'e}sz{\'a}ros, Derek~B. Fox, Chad Hanna, and Kohta Murase.
\newblock Multi-messenger astrophysics.
\newblock {\em Nature Reviews Physics}, 1(10):585--599, Oct 2019.

\bibitem{Smith03042019}
Aaron Smith and Volker~Bromm and.
\newblock Supermassive black holes in the early universe.
\newblock {\em Contemporary Physics}, 60(2):111--126, 2019.

\bibitem{Zou_2024}
Fan Zou, Zhibo Yu, W.~N. Brandt, Hyungsuk Tak, Guang Yang, and Qingling Ni.
\newblock Mapping the growth of supermassive black holes as a function of galaxy stellar mass and redshift.
\newblock {\em The Astrophysical Journal}, 964(2):183, mar 2024.

\bibitem{10.1093/mnras/stae1819}
Aklant~K Bhowmick, Laura Blecha, Paul Torrey, Rachel~S Somerville, Luke~Zoltan Kelley, Mark Vogelsberger, Rainer Weinberger, Lars Hernquist, and Aneesh Sivasankaran.
\newblock Growth of high-redshift supermassive black holes from heavy seeds in the brahma cosmological simulations: implications of overmassive black holes.
\newblock {\em Monthly Notices of the Royal Astronomical Society}, 533(2):1907--1926, 07 2024.

\bibitem{10.1093/mnras/266.1.137}
T.~R. Marsh, E.~L. Robinson, and J.~H. Wood.
\newblock Spectroscopy of \$a0620-00\$: the mass of the black hole and an image of its accretion disc.
\newblock {\em Monthly Notices of the Royal Astronomical Society}, 266(1):137--154, 01 1994.

\bibitem{10.1046/j.1365-8711.2002.05871.x}
R.~J. McLure and M.~J. Jarvis.
\newblock Measuring the black hole masses of high-redshift quasars.
\newblock {\em Monthly Notices of the Royal Astronomical Society}, 337(1):109--116, 11 2002.

\bibitem{10.1093/mnras/stt157}
G.~Calderone, G.~Ghisellini, M.~Colpi, and M.~Dotti.
\newblock Black hole mass estimate for a sample of radio-loud narrow-line seyfert 1 galaxies.
\newblock {\em Monthly Notices of the Royal Astronomical Society}, 431(1):210--239, 03 2013.

\bibitem{10.1111/j.1365-2966.2012.21074.x}
Jonathan~C. McKinney, Alexander Tchekhovskoy, and Roger~D. Blandford.
\newblock General relativistic magnetohydrodynamic simulations of magnetically choked accretion flows around black holes.
\newblock {\em Monthly Notices of the Royal Astronomical Society}, 423(4):3083--3117, 07 2012.

\bibitem{Gammie_1998}
Charles~F. Gammie and Robert Popham.
\newblock Advection-dominated accretion flows in the kerr metric. i. basic equations.
\newblock {\em The Astrophysical Journal}, 498(1):313, may 1998.

\bibitem{Popham_1998}
Robert Popham and Charles~F. Gammie.
\newblock Advection-dominated accretion flows in the kerr metric. ii. steady state global solutions.
\newblock {\em The Astrophysical Journal}, 504(1):419, sep 1998.

\bibitem{Abramowicz2013}
Marek~A. Abramowicz and P.~Chris Fragile.
\newblock Foundations of black hole accretion disk theory.
\newblock {\em Living Reviews in Relativity}, 16(1):1, Jan 2013.

\bibitem{NT1}
I.~D. {Novikov} and K.~S. {Thorne}.
\newblock {Astrophysics of black holes.}
\newblock In C.~{Dewitt} and B.~S. {Dewitt}, editors, {\em Black Holes (Les Astres Occlus)}, pages 343--450, January 1973.

\bibitem{NT2}
Don~N. {Page} and Kip~S. {Thorne}.
\newblock {Disk-Accretion onto a Black Hole. Time-Averaged Structure of Accretion Disk}.
\newblock {\em \apj}, 191:499--506, July 1974.

\bibitem{EDGAR2004843}
Richard Edgar.
\newblock A review of bondi–hoyle–lyttleton accretion.
\newblock {\em New Astronomy Reviews}, 48(10):843--859, 2004.

\bibitem{Ricotti_2007}
Massimo Ricotti.
\newblock Bondi accretion in the early universe.
\newblock {\em The Astrophysical Journal}, 662(1):53, jun 2007.

\bibitem{PhysRevD.87.044007}
Janusz Karkowski and Edward Malec.
\newblock Bondi accretion onto cosmological black holes.
\newblock {\em Phys. Rev. D}, 87:044007, Feb 2013.

\bibitem{CHAUDHARY2025170006}
Shahid Chaudhary, Shahi Rome, Muhammad~Danish Sultan, Ahmadjon Abdujabbarov, Awatef Abidi, Yousef~Mohammad Alanazi, and Abdulrahman~Bin Jumah.
\newblock Impact of extended gravity theories on accretion dynamics and observables in black hole.
\newblock {\em Annals of Physics}, 477:170006, 2025.

\bibitem{Boshkayev_2022}
K.~Boshkayev, T.~Konysbayev, Ye. Kurmanov, O.~Luongo, and D.~Malafarina.
\newblock Accretion disk luminosity for black holes surrounded by dark matter with tangential pressure.
\newblock {\em The Astrophysical Journal}, 936(2):96, sep 2022.

\bibitem{Boshkayev2020}
K~Boshkayev, A~Idrissov, O~Luongo, and D~Malafarina.
\newblock Accretion disc luminosity for black holes surrounded by dark matter.
\newblock {\em Monthly Notices of the Royal Astronomical Society}, 496(2):1115--1123, 06 2020.

\bibitem{Kurmanov_2022}
E.~Kurmanov, K.~Boshkayev, R.~Giambò, T.~Konysbayev, O.~Luongo, D.~Malafarina, and H.~Quevedo.
\newblock Accretion disk luminosity for black holes surrounded by dark matter with anisotropic pressure.
\newblock {\em The Astrophysical Journal}, 925(2):210, feb 2022.

\bibitem{PhysRevD.106.063009}
Sourabh Nampalliwar, Aristomenis~I. Yfantis, and Kostas~D. Kokkotas.
\newblock Extending grmhd for thin disks to non-kerr spacetimes.
\newblock {\em Phys. Rev. D}, 106:063009, Sep 2022.

\bibitem{Bambi:2011vc}
Cosimo Bambi and Enrico Barausse.
\newblock {The Final stages of accretion onto non-Kerr compact objects}.
\newblock {\em Phys. Rev. D}, 84:084034, 2011.

\bibitem{Boshkayev:2021chc}
Kuantay Boshkayev, Talgar Konysbayev, Ergali Kurmanov, Orlando Luongo, Daniele Malafarina, and Hernando Quevedo.
\newblock {Luminosity of accretion disks in compact objects with a quadrupole}.
\newblock {\em Phys. Rev. D}, 104(8):084009, 2021.

\bibitem{Uktamov:2024ckf}
Uktamjon Uktamov, Mirzabek Alloqulov, Sanjar Shaymatov, Tao Zhu, and Bobomurat Ahmedov.
\newblock {Particle dynamics and the accretion disk around a Self-dual Black Hole immersed in a magnetic field in Loop Quantum Gravity}.
\newblock {\em Phys. Dark Univ.}, 47:101743, 2025.

\bibitem{Tahelyani:2022uxw}
Divya Tahelyani, Ashok~B. Joshi, Dipanjan Dey, and Pankaj~S. Joshi.
\newblock {Comparing thin accretion disk properties of naked singularities and black holes}.
\newblock {\em Phys. Rev. D}, 106(4):044036, 2022.

\bibitem{Kurmanov:2025uwq}
Yergali Kurmanov, Kuantay Boshkayev, Talgar Konysbayev, Marco Muccino, Orlando Luongo, Ainur Urazalina, Anar Dalelkhankyzy, Farida Belissarova, and Madina Alimkulova.
\newblock {Accretion disk luminosity around rotating naked singularities}.
\newblock {\em Phys. Dark Univ.}, 48:101917, 2025.

\bibitem{Cemeljic:2025bqz}
Miljenko \v{C}emelji\'c, W\l{}odek Klu\'zniak, Ruchi Mishra, and Maciek Wielgus.
\newblock {Pseudo-Newtonian Simulation of a Thin Accretion Disk Around a Reissner\textendash{}Nordstr\"om Naked Singularity}.
\newblock {\em Astrophys. J.}, 981(1):69, 2025.

\bibitem{Uniyal:2023inx}
Akhil Uniyal, Sayan Chakrabarti, Reggie~C. Pantig, and Ali \"Ovg\"un.
\newblock {Nonlinearly charged black holes: Shadow and thin-accretion disk}.
\newblock {\em New Astron.}, 111:102249, 2024.

\bibitem{Sajadi:2023ybm}
Seyed~Naseh Sajadi, Mohsen Khodadi, Orlando Luongo, and Hernando Quevedo.
\newblock {Anisotropic generalized polytropic spheres: Regular 3D black holes}.
\newblock {\em Phys. Dark Univ.}, 45:101525, 2024.

\bibitem{Kurmanov:2024hpn}
Yergali Kurmanov, Kuantay Boshkayev, Talgar Konysbayev, Orlando Luongo, Nazym Saiyp, Ainur Urazalina, Gulfeiruz Ikhsan, and Gulnara Suliyeva.
\newblock {Accretion disks properties around regular black hole solutions obtained from non-linear electrodynamics}.
\newblock {\em Phys. Dark Univ.}, 46:101566, 2024.

\bibitem{10.1093/mnras/stz1698}
Hai-Nan Lin and Xin Li.
\newblock The dark matter profiles in the milky way.
\newblock {\em Monthly Notices of the Royal Astronomical Society}, 487(4):5679--5684, 06 2019.

\bibitem{PhysRevD.110.043034}
Sobhan Kazempour, Sichun Sun, and Chengye Yu.
\newblock Accreting black holes in dark matter halos.
\newblock {\em Phys. Rev. D}, 110:043034, Aug 2024.

\bibitem{Ferreira2021}
Elisa G.~M. Ferreira.
\newblock Ultra-light dark matter.
\newblock {\em The Astronomy and Astrophysics Review}, 29(1):7, Sep 2021.

\bibitem{Rogers_2023}
Keir~K. Rogers, Renée Hložek, Alex Laguë, Mikhail~M. Ivanov, Oliver~H.E. Philcox, Giovanni Cabass, Kazuyuki Akitsu, and David~J.E. Marsh.
\newblock Ultra-light axions and the s8 tension: joint constraints from the cosmic microwave background and galaxy clustering.
\newblock {\em Journal of Cosmology and Astroparticle Physics}, 2023(06):023, jun 2023.

\bibitem{PhysRevD.95.043541}
Lam Hui, Jeremiah~P. Ostriker, Scott Tremaine, and Edward Witten.
\newblock Ultralight scalars as cosmological dark matter.
\newblock {\em Phys. Rev. D}, 95:043541, Feb 2017.

\bibitem{PhysRevD.99.063015}
Maksym Deliyergiyev, Antonino Del~Popolo, Laura Tolos, Morgan Le~Delliou, Xiguo Lee, and Fiorella Burgio.
\newblock Dark compact objects: An extensive overview.
\newblock {\em Phys. Rev. D}, 99:063015, Mar 2019.

\bibitem{Brandt_2016}
Timothy~D. Brandt.
\newblock Constraints on macho dark matter from compact stellar systems in ultra-faint dwarf galaxies.
\newblock {\em The Astrophysical Journal Letters}, 824(2):L31, jun 2016.

\bibitem{Green_2021}
Anne~M Green and Bradley~J Kavanagh.
\newblock Primordial black holes as a dark matter candidate.
\newblock {\em Journal of Physics G: Nuclear and Particle Physics}, 48(4):043001, feb 2021.

\bibitem{Luongo:2025iqq}
Orlando Luongo.
\newblock {Gravitational metamaterials from optical properties of spacetime media}.
\newblock 4 2025.

\bibitem{universe6120234}
Torsten Asselmeyer-Maluga and Jerzy Król.
\newblock Dark matter as gravitational solitons in the weak field limit.
\newblock {\em Universe}, 6(12), 2020.

\bibitem{universe5100213}
Stefano Profumo, Leonardo Giani, and Oliver~F. Piattella.
\newblock An introduction to particle dark matter.
\newblock {\em Universe}, 5(10), 2019.

\bibitem{Zavala:2019gpq}
Jes\'us Zavala and Carlos~S. Frenk.
\newblock {Dark matter haloes and subhaloes}.
\newblock {\em Galaxies}, 7(4):81, 2019.

\bibitem{3Flav}
Massimo Blasone, Antonio Capolupo, and Giuseppe Vitiello.
\newblock Quantum field theory of three flavor neutrino mixing and oscillations with $\mathrm{CP}$ violation.
\newblock {\em Phys. Rev. D}, 66:025033, Jul 2002.

\bibitem{Capo2016}
Antonio Capolupo.
\newblock Dark matter and dark energy induced by condensates.
\newblock {\em Advances in High Energy Physics}, 2016(1):8089142, 2016.

\bibitem{CosmoFlav}
Antonio Capolupo, Sante Carloni, and Aniello Quaranta.
\newblock Quantum flavor vacuum in the expanding universe: A possible candidate for cosmological dark matter?
\newblock {\em Phys. Rev. D}, 105:105013, May 2022.

\bibitem{CurvFlav}
Antonio Capolupo, Aniello Quaranta, and Raoul Serao.
\newblock Field mixing in curved spacetime and dark matter.
\newblock {\em Symmetry}, 15(4), 2023.

\bibitem{Capolupo2025}
Antonio Capolupo, Salvatore Capozziello, Gabriele Pisacane, and Aniello Quaranta.
\newblock Missing matter in galaxies as a neutrino mixing effect.
\newblock {\em Physics of the Dark Universe}, 48:101894, 2025.

\bibitem{CurvNeut}
A.~Capolupo, G.~Lambiase, and A.~Quaranta.
\newblock Neutrinos in curved spacetime: Particle mixing and flavor oscillations.
\newblock {\em Phys. Rev. D}, 101:095022, May 2020.

\bibitem{ParticleDataGroup:2024cfk}
S.~Navas et~al.
\newblock {Review of particle physics}.
\newblock {\em Phys. Rev. D}, 110(3):030001, 2024.

\bibitem{Luongo:2024opv}
Orlando Luongo and Tommaso Mengoni.
\newblock {Generalized K-essence inflation in Jordan and Einstein frames}.
\newblock {\em Class. Quant. Grav.}, 41(10):105006, 2024.

\bibitem{DAgostino:2022fcx}
Rocco D'Agostino, Orlando Luongo, and Marco Muccino.
\newblock {Healing the cosmological constant problem during inflation through a unified quasi-quintessence matter field}.
\newblock {\em Class. Quant. Grav.}, 39(19):195014, 2022.

\bibitem{Belfiglio:2023rxb}
Alessio Belfiglio, Youri Carloni, and Orlando Luongo.
\newblock {Particle production from non-minimal coupling in a symmetry breaking potential transporting vacuum energy}.
\newblock {\em Phys. Dark Univ.}, 44:101458, 2024.

\bibitem{Luongo:2018lgy}
Orlando Luongo and Marco Muccino.
\newblock {Speeding up the universe using dust with pressure}.
\newblock {\em Phys. Rev. D}, 98(10):103520, 2018.

\bibitem{Belfiglio:2022qai}
Alessio Belfiglio, Roberto Giamb{\`o}, and Orlando Luongo.
\newblock {Alleviating the cosmological constant problem from particle production}.
\newblock {\em Class. Quant. Grav.}, 40(10):105004, 2023.

\bibitem{Luongo:2023jnb}
Orlando Luongo.
\newblock {Revising the cosmological constant problem through a fluid different from the quintessence}.
\newblock {\em Phys. Sci. Tech.}, 10(3-4):17--27, 2023.

\bibitem{10.1093/mnras/stab2571}
K~Boshkayev, T~Konysbayev, E~Kurmanov, O~Luongo, D~Malafarina, K~Mutalipova, and G~Zhumakhanova.
\newblock Effects of non-vanishing dark matter pressure in the milky way galaxy.
\newblock {\em Monthly Notices of the Royal Astronomical Society}, 508(1):1543--1554, 09 2021.

\bibitem{Inan_2024}
Nader Inan, Ahmed~Farag Ali, Kimet Jusufi, and Abdelrahman Yasser.
\newblock Graviton mass due to dark energy as a superconducting medium-theoretical and phenomenological aspects.
\newblock {\em Journal of Cosmology and Astroparticle Physics}, 2024(08):012, aug 2024.

\bibitem{1984A&A...136L..21S}
R.~H. {Sanders}.
\newblock {Anti-gravity and galaxy rotation curves}.
\newblock {\em aap}, 136(2):L21--L23, July 1984.

\bibitem{CAPOZZIELLO2020100573}
Salvatore Capozziello, Vesna~Borka Jovanović, Duško Borka, and Predrag Jovanović.
\newblock Constraining theories of gravity by fundamental plane of elliptical galaxies.
\newblock {\em Physics of the Dark Universe}, 29:100573, 2020.

\bibitem{2007MNRAS.381.1103F}
C.~{Frigerio Martins} and P.~{Salucci}.
\newblock {Analysis of rotation curves in the framework of R$^{n}$ gravity}.
\newblock {\em MNRAS}, 381(3):1103--1108, November 2007.

\bibitem{PhysRevD.104.043009}
Jakob Henrichs, Margherita Lembo, Fabio Iocco, and Luca Amendola.
\newblock Testing gravity with the milky way: Yukawa potential.
\newblock {\em Phys. Rev. D}, 104:043009, Aug 2021.

\bibitem{Faber:2005xc}
Tristan Faber and Matt Visser.
\newblock {Combining rotation curves and gravitational lensing: How to measure the equation of state of dark matter in the galactic halo}.
\newblock {\em Mon. Not. Roy. Astron. Soc.}, 372:136--142, 2006.

\bibitem{Harko2009}
{Kovács, Z.}, {Cheng, K. S.}, and {Harko, T.}
\newblock Thin accretion discs around neutron and quark stars.
\newblock {\em A\&A}, 500(2):621--631, 2009.

\end{thebibliography}
\end{document}